\documentclass[titlepage,12pt]{article}
\usepackage{amssymb}
\usepackage{amsfonts}
\usepackage{amsmath}
\usepackage[dvips]{graphicx}
\usepackage[font=small,format=plain,labelfont=bf,up,textfont=it,up]{caption}

\begin{document}

\title{Remarks on the spin-one Duffin-Kemmer-Petiau equation in the presence
of nonminimal vector interactions in (3+1) dimensions}
\author{L. B. Castro\thanks{%
luis.castro@pgfsc.ufsc.br} \\
Departamento de F\'{\i}sica, CFM\\
Universidade Federal de Santa Catarina (UFSC),\\
88040-900, CP. 476, Florian\'{o}polis, SC, Brazil\\
and\\
Departamento de F\'{\i}sica, \\
Universidade Federal do Maranh\~{a}o (UFMA), \\
Campus Universit\'{a}rio do Bacanga, 65085-580, S\~{a}o Lu\'{\i}s, MA,
Brazil \\
and\\
L. P. de Oliveira\thanks{%
luizp@if.usp.br}\\
Instituto de F\'{\i}sica,\\
Universidade de S\~{a}o Paulo (USP),\\
05508-900, S\~{a}o Paulo, SP, Brazil}
\date{}
\maketitle

\begin{abstract}
We point out a misleading treatment in the recent literature regarding
analytical solutions for nonminimal vector interaction for spin-one
particles in the context of the Duffin-Kemmer-Petiau (DKP) formalism. In
those papers, the authors use improperly the nonminimal vector interaction
endangering in their main conclusions. We present a few properties of the
nonminimal vector interactions and also present the correct equations to
this problem. We show that the solution can be easily found by solving Schr%
\"{o}dinger-like equations. As an application of this procedure, we consider
spin-one particles in presence of a nonminimal vector linear potential.

\bigskip


\bigskip

\end{abstract}

\section{Introduction}

The Duffin-Kemmer-Petiau (DKP) formalism \cite{pet}-\cite{kem} describes
spin-zero and spin-one particles and has been used to analyze relativistic
interactions of spin-zero and spin-one hadrons with nuclei as an alternative
to their conventional second-order Klein-Gordon (KG) and Proca counterparts.
The DKP formalism proved to be better that the KG formalism in the analysis
of $K_{l3}$ decays, the decay-rate ratio $\Gamma (\eta \rightarrow \gamma
\gamma )/\Gamma (\pi ^{0}\rightarrow \gamma \gamma )$, and level shifts and
widths in pionic atoms~\cite{fis1}-\cite{parte 11}. The DKP formalism enjoys
a richness of couplings not capable of being expressed in the KG and Proca
theories \cite{gue}-\cite{vij}. Although the formalisms are equivalent in
the case of minimally coupled vector interactions \cite{mr}-\cite{lun}, the
DKP formalism opens new horizons as far as it allows other kinds of
couplings which are not possible in the KG and Proca theories. The
nonminimal vector interaction refers to a kind of charge conjugate invariant
coupling that behaves like a vector under a Lorentz transformation. The
invariance of the nonminimal vector potential under charge conjugation means
that it does not distinguish particles from antiparticles. Hence, whether
one consider spin-zero or spin-one bosons, this sort of interaction cannot
exhibit Klein\`{}s paradox \cite{jpa}. Nonminimal vector potentials, added
by other kinds of Lorentz structures, have already been used in a
phenomenological context for describing the scattering of mesons by nuclei
\cite{cla1}-\cite{cla2}, but it should be mentioned that in Refs. \cite{cla1}%
-\cite{koz2}, \cite{kur1}-\cite{kur}, \cite{cla2} the nonminimal vector
couplings have been used improperly. Nonminimal vector coupling with a
quadratic potential \cite{Ait}, with a linear potential \cite{kuli}, and
mixed space and time components with a step potential \cite{ccc3}-\cite{ccc2}%
, double-step potential \cite{lp}, a smooth step potential \cite{ben3}, a
linear potential \cite{jpa}, \cite{ben4}, \cite{ben}, and a linear plus
inversely linear potential \cite{ben2}, have been explored in the
literature. In a recent paper published in this journal, Hassanabadi and
collaborators \cite{erro} analyze the DKP equation in the presence of
nonminimal vectorial interactions (Coulomb and harmonic oscillator
potentials) in (3+1) dimensions for spin-one particles. In that paper, the
authors used improperly the nonminimal vector interaction endangering its
main conclusions. The same mistake is found in recent works \cite{erro1}-%
\cite{erro9}, for instance. Other misconception is found in Refs. \cite{cla1}%
-\cite{koz2}, in where the space component of the nonminimal vector
potential is absorbed into the spinor. As it is shown in \cite{jpa}, there
is no chance to dissociate from this term. Furthermore, the space component
of the nonminimal vector potential could be irrelevant for the formation of
bound states for potentials vanishing at infinity, but its presence is an
essential ingredient for confinement.

In view of the misconceptions on the nonminimal vector interaction
propagated in the literature, the purpose of this Review Article is to
review the DKP equation in the presence of a nonminimal vectorial
interaction for spin-one particles in (3+1) dimensions. We present a few
properties of the nonminimal vector interactions and also present the
correct equations to this problem. We show that the solution can be easily
found by solving Schr\"{o}dinger-like equations. As an application of this
procedure, we consider spin-one particles in presence of a nonminimal vector
linear potential. For this case in particular, the problem is mapped into
the nonrelativistic three-dimensional harmonic oscillator.

\section{The DKP equation}

The DKP equation for a free boson is given by \cite{kem} (with units in
which $\hbar =c=1$)%
\begin{equation}
\left( i\beta ^{\mu }\partial _{\mu }-m\right) \psi =0  \label{dkp}
\end{equation}%
\noindent where the matrices $\beta ^{\mu }$\ satisfy the algebra
\begin{equation}  \label{betaalge}
\beta^{\mu }\beta ^{\nu }\beta ^{\lambda }+\beta ^{\lambda }\beta ^{\nu
}\beta ^{\mu }=g^{\mu \nu }\beta ^{\lambda }+g^{\lambda \nu }\beta ^{\mu }
\end{equation}

\noindent and the metric tensor is $g^{\mu \nu }=\,$diag$\,(1,-1,-1,-1)$.
The algebra expressed by (\ref{betaalge}) generates a set of 126 independent
matrices whose irreducible representations are a trivial representation, a
five-dimensional representation describing the spin-zero particles and a
ten-dimensional representation associated to spin-one particles. The DKP
spinor has an excess of components and the theory has to be supplemented by
an equation which allows us eliminate the redundant components. That
constraint equation is obtained by multiplying the DKP equation by $1-\beta
^{0}\beta ^{0}$, namely%
\begin{equation}
i\beta ^{i}\beta ^{0}\beta ^{0}\partial _{i}\psi =m\left( 1-\beta ^{0}\beta
^{0}\right) \psi  \label{vin1}
\end{equation}

\noindent This constraint equation expresses three (four) components of the
spinor by the other two (six) components and their space derivatives in the
scalar (vector) sector so that the superfluous components disappear and
there only remain the physical components of the DKP theory. The
second-order Klein-Gordon and Proca equations are obtained when one selects
the spin-zero and spin-one sectors of the DKP theory.

A well-known conserved four-current is given by
\begin{equation}  \label{cu}
J^{\mu }=\frac{1}{2}\,\bar{\psi}\beta ^{\mu }\psi
\end{equation}

\noindent where the adjoint spinor $\bar{\psi}$ is given by $\bar{\psi}=\psi
^{\dagger }\eta ^{0}$ with $\eta ^{0}=2\beta ^{0}\beta ^{0}-1$ in such a way
that $(\eta^{0}\beta^{\mu})^{\dag}=\eta^{0}\beta^{\mu}$ (the matrices $%
\beta^{\mu}$ are Hermitian with respect to $\eta^{0}$). Despite the
similarity to the Dirac equation, the DKP equation involves singular
matrices, the time component of $J^{\mu}$ is not
positive definite and the case of massless bosons cannot be obtained by a
limiting process~\cite{ben6}. Nevertheless, the matrices $\beta^{\mu}$ plus
the unit operator generate a ring consistent with integer-spin algebra and $%
J^{0}$ may be interpreted as a charge density. The normalization condition $%
\int d\tau \,J^{0}=\pm 1$ can be expressed as%
\begin{equation}
\int d\tau \,\bar{\psi}\beta ^{0}\psi =\pm 2  \label{norm}
\end{equation}

\noindent where the plus (minus) sign must be used for a positive (negative)
charge.

\section{Interactions in the DKP equation}

With the introduction of interactions, the DKP equation can be written as%
\begin{equation}
\left( i\beta ^{\mu }\partial _{\mu }-m-U\right) \psi =0  \label{dkp22}
\end{equation}

\noindent where the more general potential matrix $U$ is written in terms of
25 (100) linearly independent matrices pertinent to five (ten)-dimensional
irreducible representation associated to the scalar (vector) sector. In the
presence of interaction, $J^{\mu}$ satisfies the equation
\begin{equation}  \label{corrent2}
\partial _{\mu }J^{\mu }+\frac{i}{2}\,\bar{\psi}\left( U-\eta ^{0}U^{\dagger
}\eta ^{0}\right) \psi =0
\end{equation}

\noindent Thus, if $U$ is Hermitian with respect to $\eta ^{0}$ then
four-current will be conserved. The potential matrix $U$ can be written in
terms of well-defined Lorentz structures. For the spin-zero sector there are
two scalar, two vector and two tensor terms \cite{gue}, whereas for the
spin-one sector there are two scalar, two vector, a pseudoscalar, two
pseudovector and eight tensor terms \cite{vij}. The tensor terms have been
avoided in applications because they furnish noncausal effects \cite{gue}-%
\cite{vij}.

\subsection{Nonminimal vector couplings in the DKP equation}

Considering only the nonminimal vector interaction, the DKP equation can be
written as
\begin{equation}
\left( i\beta ^{\mu }\partial _{\mu }-m-i[P,\beta ^{\mu }]A_{\mu }\right)
\psi =0  \label{dkp2}
\end{equation}

\noindent where $P$ is a projection operator ($P^{2}=P$ and $P^{\dagger }=P$%
) in such a way that $\bar{\psi}[P,\beta ^{\mu }]\psi $ behaves like a
vector under a Lorentz transformation as does $\bar{\psi}\beta ^{\mu }\psi $%
. One very important point to note is that this potential leads to a
conserved four-current but the same does not happen if instead of $%
i[P,\beta^{\mu}]$ one uses either $P\beta^{\mu}$ or $\beta^{\mu}P$, as in
\cite{cla1}-\cite{koz2}, \cite{kur1}-\cite{kur}, \cite{cla2}, \cite{erro}-%
\cite{erro9}. As a matter of fact, in \cite{cla1} it is mentioned that $%
P\beta^{\mu}$ and $\beta^{\mu}P$ produce identical results. Considering
explicitly the condition (\ref{corrent2}) for the potential $%
U=\beta^{0}PV_{p}$ (widely used in the literature), we obtain
\begin{equation}  \label{curr}
\partial_{\mu}J^{\mu}=\frac{i}{2}\bar{\psi}[P,\beta^{0}]V_{p}\psi\neq0\,.
\end{equation}

\noindent The current is not conserved and it is proportional to $V_{p}$.
The fact that this current is not conserved has crucial consequences on the orthonormal
condition of the DKP spinor \cite{jpa,ccc3,ccc2,ben3,ben4}.

The DKP equation is invariant under the parity operation, i.e. when $\vec{r}%
\rightarrow-\vec{r}$, if $\vec{A}$ changes sign, whereas $A_{0}$ remains the
same. This is because the parity operator is $\mathcal{P}=\mathrm{exp}%
(i\delta_{p})P_{0}\eta^{0}$, where $\delta_{p}$ is a constant phase and $%
P_{0}$ changes $\vec{r}$ into $-\vec{r}$. Because this unitary operator
anticommutes with $\beta^{i}$ and $[P,\beta^{i}]$, they change the sign
under a parity transformation, whereas $\beta^{0}$ and $[P,\beta^{0}]$,
which commute with $\eta^{0}$, remain the same. Since $\delta_{p}=0$ or $%
\delta_{p}=\pi$, the spinor components have definite parities. The charge
conjugation operation can be accomplished by the transformation $%
\psi\rightarrow\psi_{c}=\mathcal{C}\psi=CK\psi$, where $K$ denotes the
complex conjugation and $C$ is a unitary matrix such that $%
C\beta^{\mu}=-\beta^{\mu}C$. The matrix that satisfies these relations is $C=%
\mathrm{exp}(i\delta_{C})\eta^{0}\eta^{1}$. The phase factor $\mathrm{exp}%
(i\delta_{C})$ is equal to $\pm1$, thus $E\rightarrow-E$. Note also that $%
J^{\mu}\rightarrow-J^{\mu}$, as should be expected for a charge current.
Meanwhile $C$ anticommutes with $[P,\beta^{\mu}]$ and the charge conjugation
operation entails no change on $A_{\mu}$. The invariance of the nonminimal
vector potential under charge conjugation means that it does not couple to
the charge of the boson. In other words, $A_{\mu}$ does not distinguish
particles from antiparticles. Hence, whether one considers spin-zero or
spin-one bosons, this sort of interaction cannot exhibit Klein\`{}s paradox~%
\cite{jpa}.

If the potential is time-independent one can write $\psi (\vec{r},t)=\varphi
(\vec{r})\exp (-iEt)$, where $E$ is the energy of the boson, in such a way
that the time-independent DKP equation becomes%
\begin{equation}
\left[ \beta ^{0}E+i\beta ^{i}\partial _{i}-\left( m+i[P,\beta ^{\mu
}]A_{\mu }\right) \right] \varphi =0  \label{DKP10}
\end{equation}%
In this case \ $J^{\mu }=\bar{\varphi}\beta ^{\mu }\varphi /2$ does not
depend on time, so that the spinor $\varphi $ describes a stationary state.

\subsection{Vectorial sector}

For the case of spin-one (vectorial sector), the $\beta ^{\mu }$\ matrices
are~\cite{ben5}%
\begin{equation}
\beta ^{0}=%
\begin{pmatrix}
0 & \overline{0} & \overline{0} & \overline{0} \\
\overline{0}^{T} & \mathbf{0} & \mathbf{I} & \mathbf{0} \\
\overline{0}^{T} & \mathbf{I} & \mathbf{0} & \mathbf{0} \\
\overline{0}^{T} & \mathbf{0} & \mathbf{0} & \mathbf{0}%
\end{pmatrix}%
,\quad \beta ^{i}=%
\begin{pmatrix}
0 & \overline{0} & e_{i} & \overline{0} \\
\overline{0}^{T} & \mathbf{0} & \mathbf{0} & -is_{i} \\
-e_{i}^{T} & \mathbf{0} & \mathbf{0} & \mathbf{0} \\
\overline{0}^{T} & -is_{i} & \mathbf{0} & \mathbf{0}%
\end{pmatrix}
\label{betaspin1}
\end{equation}

\noindent where $s_{i}$ are the 3$\times $3 spin-1 matrices $\left(
s_{i}\right) _{jk}=-i\varepsilon _{ijk}$, $e_{i}$ are the 1$\times $3
matrices $\left( e_{i}\right) _{1j}=\delta _{ij}$ and $\overline{0}=%
\begin{pmatrix}
0 & 0 & 0%
\end{pmatrix}%
$, while\textbf{\ }$\mathbf{I}$ and $\mathbf{0}$\textbf{\ }designate the 3$%
\times $3 unit and zero matrices, respectively, while the superscript T
designates matrix transposition. In this representation $P=\,\beta ^{\mu
}\beta _{\mu }-2=\mathrm{diag}\,(1,1,1,1,0,0,0,0,0,0)$, i.e. $P$ projects
out the four upper components of the DKP spinor. The ten-component spinor
can be written as $\varphi ^{T}=\left( \varphi _{1},...,\varphi _{10}\right)$
and partitioned as (following the notation of Ref.~\cite{erro})
\begin{equation}
\varphi_{1}= i\phi,\quad \vec{F}=\left(
\begin{array}{c}
\varphi_{2} \\
\varphi_{3} \\
\varphi_{4}%
\end{array}%
\right)
\end{equation}%
\begin{equation}
\vec{G}=\left(
\begin{array}{c}
\varphi_{5} \\
\varphi_{6} \\
\varphi_{7}%
\end{array}%
\right) ,\quad \vec{H}=\left(
\begin{array}{c}
\varphi_{8} \\
\varphi_{9} \\
\varphi_{10}%
\end{array}%
\right)  \label{part}
\end{equation}

\noindent the DKP equation in (3+1) dimensions can be expressed in the
compact form
\begin{eqnarray}  \label{eqms1}
i\vec{\nabla}\times\vec{F}-i \vec{A}\times\vec{F} &=& m\vec{H}  \label{eqms1}
\\
\vec{\nabla}\cdot\vec{G}+ \vec{A}\cdot\vec{G} &=& m\phi  \label{eqms2} \\
i\vec{\nabla}\times\vec{H}+i \vec{A}\times\vec{H} &=& m\vec{F}-(E-iA_{0})%
\vec{G}  \label{eqms3} \\
\vec{\nabla}\phi- \vec{A}\phi &=& m\vec{G}-(E+iA_{0})\vec{F}  \label{eqms4}
\end{eqnarray}

\noindent At this stage is worthwhile to mention that the eqs.~(\ref{eqms1}%
)-(\ref{eqms4}) are completely different from those given in \cite{erro} and
this fact is due to use improperly the nonminimal vector coupling. These
facts should be enough to jeopardize the results presented in the Refs.~\cite%
{erro}-\cite{erro9}.

Using the standard procedure developed in \cite{neds1}, we put
\begin{equation}  \label{eqc1}
\phi=\frac{\phi_{n\,j}(r)}{r}Y_{j}^{m_{j}}(\theta,\varphi)
\end{equation}
\begin{equation}  \label{eqc2}
\vec{F}=\sum_{l}\frac{F_{n\,j\,l}(r)}{r}\vec{Y}_{j\,l\,m_{j}}(\theta,\varphi)
\end{equation}
\begin{equation}  \label{eqc3}
\vec{G}=\sum_{l}\frac{G_{n\,j\,l}(r)}{r}\vec{Y}_{j\,l\,m_{j}}(\theta,\varphi)
\end{equation}
\begin{equation}  \label{eqc4}
\vec{H}=\sum_{l}\frac{H_{n\,j\,l}(r)}{r}\vec{Y}_{j\,l\,m_{j}}(\theta,\varphi)
\end{equation}

\noindent where $\phi_{n\,j}$, $F_{n\,j\,l}$, $G_{n\,j\,l}$ and $H_{n\,j\,l}$
are radial wave functions while $Y_{j}^{m_{j}}(\theta,\varphi)$ are the
usual spherical harmonics of order $j$, and $\vec{Y}_{j\,l\,m_{j}}(\theta,%
\varphi)$ are the vector spherical harmonics. Then, using the notation
\begin{equation}  \label{nota}
F_{n\,j\,j}=F_{0}, \qquad F_{n\,j\,j\pm1}=F_{\pm1}
\end{equation}

\noindent and similar definitions for $G_{0}$, $G_{\pm1}$, $H_{0}$ and $%
H_{\pm1}$ together with the properties of vector spherical harmonics (see
Appendix A), we can get a set of first-order coupled differential radial
equations. Substituting (\ref{eqc2}) and (\ref{eqc4}) in (\ref{eqms1}) and
if we consider spherically symmetric potentials $A_{0}=A_{0}(r)$ and $\vec{A}%
=A_{r}(r)\hat{r}$, the radial differential equations obtained from (\ref%
{eqms1}) are
\begin{equation}  \label{eqr1}
\left( \frac{dF_{0}}{dr}-\frac{j+1}{r}F_{0}-A_{r}F_{0} \right)=-\frac{1}{%
\zeta_{j}}mH_{+1}
\end{equation}
\begin{equation}  \label{eqr2}
\left( \frac{dF_{0}}{dr}+\frac{j}{r}F_{0}-A_{r}F_{0} \right)=-\frac{1}{%
\alpha_{j}}mH_{-1}
\end{equation}
\begin{equation}  \label{eqr3}
-\zeta_{j}\left( \frac{dF_{+1}}{dr}+\frac{j+1}{r}F_{+1}-A_{r}F_{+1} \right)-
\alpha_{j}\left( \frac{dF_{-1}}{dr}-\frac{j}{r}F_{-1}-A_{r}F_{-1}%
\right)=mH_{0}
\end{equation}

\noindent where $\alpha_{j}=\sqrt{(j+1)/(2j+1)}$ and $\zeta_{j}=\sqrt{%
j/(2j+1)}$.

\noindent Similarly, substituting (\ref{eqc1}) and (\ref{eqc3}) in (\ref%
{eqms2}), we obtain
\begin{equation}  \label{eqr4}
-\alpha_{j}\left( \frac{dG_{+1}}{dr}+\frac{j+1}{r}G_{+1}+A_{r}G_{+1}
\right)+ \zeta_{j}\left( \frac{dG_{-1}}{dr}-\frac{j}{r}G_{-1}+A_{r}G_{-1}%
\right)=m\phi
\end{equation}

\noindent The radial equations obtained from (\ref{eqms3}) are
\begin{equation}  \label{eqr5}
\left( \frac{dH_{0}}{dr}-\frac{j+1}{r}H_{0}+A_{r}H_{0} \right)=-\frac{1}{%
\zeta_{j}}\left( mF_{+1}-(E-iA_{0})G_{+1} \right)
\end{equation}
\begin{equation}  \label{eqr6}
\left( \frac{dH_{0}}{dr}+\frac{j}{r}H_{0}+A_{r}H_{0} \right)=-\frac{1}{%
\alpha_{j}}\left( mF_{-1}-(E-iA_{0})G_{-1} \right)
\end{equation}
\begin{eqnarray}  \label{eqr7}
-\zeta_{j}\left( \frac{dH_{+1}}{dr}+\frac{j+1}{r}H_{+1}+A_{r}H_{+1} \right)-
& &  \notag \\
\alpha_{j}\left( \frac{dH_{-1}}{dr}-\frac{j}{r}H_{-1}+A_{r}H_{-1}\right) &=&
\left( mF_{0}-(E-iA_{0})G_{0} \right)
\end{eqnarray}

\noindent Finally, from (\ref{eqms4}) we get
\begin{equation}  \label{eqr8}
\left( E+iA_{0} \right)F_{0}=mG_{0}
\end{equation}
\begin{equation}  \label{eqr9}
\left( \frac{d\phi}{dr}-\frac{j+1}{r}\phi-A_{r}\phi \right)=-\frac{1}{%
\alpha_{j}}\left( mG_{+1}-(E+iA_{0})F_{+1} \right)
\end{equation}
\begin{equation}  \label{eqr10}
\left( \frac{d\phi}{dr}+\frac{j}{r}\phi-A_{r}\phi \right)=\frac{1}{\zeta_{j}}%
\left( mG_{-1}-(E+iA_{0})F_{-1} \right)
\end{equation}

Now, following the procedure used in Ref.~\cite{neds1}, the ten coupled
differential radial equations obtained above (equations (\ref{eqr1})-(\ref%
{eqr10})) can be decoupled into two classes of radial equations associated
to a specific parity. For states of $(-1)^{j}$ parity, the relevant
differential equations are the eqs. (\ref{eqr1}), (\ref{eqr2}), (\ref{eqr7})
and (\ref{eqr8}). The remaining six radial wave functions are zero. On the
other hand, for states of $(-1)^{j+1}$ parity, the relevant differential
equations are the eqs. (\ref{eqr3}), (\ref{eqr4}), (\ref{eqr5}), (\ref{eqr6}%
), (\ref{eqr9}) and (\ref{eqr10}). Similarly to the previous case, the other
four radial wave functions are zero.

\subsubsection{$(-1)^{j}$ parity states}

Using the eqs. (\ref{eqr1}), (\ref{eqr2}) and (\ref{eqr8}) the components $%
H_{+1}$, $H_{-1}$ and $G_{0}$ can be eliminated in favor of $F_{0}$ then by
inserting they in (\ref{eqr7}), the radial function $F_{0}(r)$ obeys the
second-order differential equation
\begin{equation}  \label{schro1}
\frac{d^{2}F_{0}(r)}{dr^{2}}+\left[ \kappa^{2}-\frac{dA_{r}}{dr}-\frac{j(j+1)}{r^{2}%
}-A_{r}^{2} \right]F_{0}(r)=0
\end{equation}

\noindent where $\kappa^{2}=E^{2}-m^{2}+A_{0}^{2}$ and because $%
\nabla^{2}(1/r)=-4\pi\delta(\vec{r})$, unless the potentials contain a delta
function at the origin, one must impose the homogeneous Dirichlet condition $%
F_{0}(0)=0$. At this stage is worthwhile to mention that (\ref{schro1}) is
very similar to DKP equation for spin-zero particles in (3+1) dimensions
except for the term $-2A_{r}/r$ \cite{ben}. Therefore, for motion in a
central field, the solution of the three-dimensional DKP equation with
nonminimal vectorial interaction can be found by solving a Schr\"{o}%
dinger-like equation for states of $(-1)^{j}$ parity. The other components
are obtained through of (\ref{eqr1}), (\ref{eqr2}) and (\ref{eqr8}).

\subsubsection{$(-1)^{j+1}$ parity states}
\label{sec322}

Using the eqs. (\ref{eqr5}) and (\ref{eqr9}), we obtain
\begin{equation}  \label{eeqq1}
\left(%
\begin{array}{c}
F_{+1} \\
G_{+1}%
\end{array}%
\right) =\frac{1}{\kappa^{2}} \left(
\begin{array}{cc}
(E-iA_{0})\alpha_{j}\Delta_{-} & m\zeta_{j}\Delta_{+} \\
m\alpha_{j}\Delta_{-} & (E+iA_{0})\zeta_{j}\Delta_{+}%
\end{array}
\right) \left(
\begin{array}{c}
\phi \\
H_{0}%
\end{array}
\right)\,,
\end{equation}

\noindent where $\Delta_{\pm}=\frac{d}{dr}-\frac{j+1}{r}\pm A_{r}$.
Similarly, using (\ref{eqr6}) and (\ref{eqr10}), we get
\begin{equation}  \label{eeqq2}
\left(%
\begin{array}{c}
F_{-1} \\
G_{-1}%
\end{array}%
\right) =\frac{1}{\kappa^{2}} \left(
\begin{array}{cc}
-(E-iA_{0})\zeta_{j}\Xi_{-} & m\alpha_{j}\Xi_{+} \\
-m\zeta_{j}\Xi_{-} & (E+iA_{0})\alpha_{j}\Xi_{+}%
\end{array}
\right) \left(
\begin{array}{c}
\phi \\
H_{0}%
\end{array}
\right)\,,
\end{equation}

\noindent where $\Xi_{\pm}=\frac{d}{dr}+\frac{j}{r}\pm A_{r}$. In this
general case, we are not able to obtain analytical solutions to this kind
of parity states, because we can not decouple the differential equations for
the components $H_{0}$ and $\phi$. An alternative to overcome this
disadvantage is to restrict our analysis for $j=0$. Considering $j=0$, we
get decouple the differential equations for the components $H_{0}$ and $\phi$%
, but those differential equations are very complicated and do not furnish
exact solutions.

For $A_{0}=0$, the equations (\ref{eeqq1}) and (\ref{eeqq2}) reduce to
\begin{equation}  \label{eeqq3}
\left(%
\begin{array}{c}
F_{+1} \\
G_{+1}%
\end{array}%
\right) =\frac{1}{\bar{\kappa}^{2}} \left(
\begin{array}{cc}
E\alpha_{j}\Delta_{-} & m\zeta_{j}\Delta_{+} \\
m\alpha_{j}\Delta_{-} & E\zeta_{j}\Delta_{+}%
\end{array}
\right) \left(
\begin{array}{c}
\phi \\
H_{0}%
\end{array}
\right)\,,
\end{equation}

\noindent and
\begin{equation}  \label{eeqq4}
\left(%
\begin{array}{c}
F_{-1} \\
G_{-1}%
\end{array}%
\right) =\frac{1}{\bar{\kappa}^{2}} \left(
\begin{array}{cc}
-E\zeta_{j}\Xi_{-} & m\alpha_{j}\Xi_{+} \\
-m\zeta_{j}\Xi_{-} & E\alpha_{j}\Xi_{+}%
\end{array}
\right) \left(
\begin{array}{c}
\phi \\
H_{0}%
\end{array}
\right)\,,
\end{equation}

\noindent where $\bar{\kappa}^{2}=E^{2}-m^{2}$. In this case, we obtain that
the radial functions $H_{0}(r)$ and $\phi(r)$ obey the second-order
differential equations
\begin{equation}  \label{schro2}
\frac{d^{2}H_{0}}{dr^{2}}+\left[ \bar{\kappa}^{2}+\frac{dA_{r}}{dr}-\frac{%
j(j+1)}{r^{2}}-A_{r}^{2} \right]H_{0}=0\,,
\end{equation}
\begin{equation}  \label{schro3}
\frac{d^{2}\phi}{dr^{2}}+\left[ \bar{\kappa}^{2}-\frac{dA_{r}}{dr}-\frac{%
j(j+1)}{r^{2}}-\frac{A_{r}}{r}-A_{r}^{2} \right]\phi=0\,.
\end{equation}

\noindent Therefore, for the particular case $A_{0}(r)=0$, the
solution of the three-dimensional DKP equation with nonminimal vectorial
interaction can be found by solving two Schr\"{o}dinger-like equations for
states of $(-1)^{j+1}$ parity. The other components are obtained through of (%
\ref{eeqq3}) and (\ref{eeqq4}). It should not be forgotten, though, that the
equations for $H_{0}$ and $\phi$ are not indeed independent because the
energy $E$ appears in both equations. Therefore, one has to search for bound-state solutions for $H_{0}$ and $\phi$ with a common energy.

\subsection{Nonminimal vector linear potential}

Having set up the spin-one equations for nonminimal vector interaction, we
are now in a position to use the machinery developed above in order to solve
the DKP equation for some specific form of the nonminimal interaction. As an
application of this procedure, let us consider a nonminimal vector linear
potential in the form
\begin{equation}  \label{potl}
A_{0}=m^2\lambda_{0}r\, \qquad A_{r}=m^2\lambda_{r}r
\end{equation}

\noindent where $\lambda_{0}$ and $\lambda_{r}$ are dimensionless
quantities.

\subsubsection{$(-1)^{j}$ parity states}

Substituting (\ref{potl}) in (\ref{schro1}), one finds that $%
F_{0}(r) $ obeys the second-order differential equation
\begin{equation}  \label{eqoh}
\frac{d^{2}F_{0}}{dr^{2}}+\left[ K^{2}-\lambda^{2}r^{2}-\frac{j(j+1)}{r^{2}} %
\right]F_{0}=0
\end{equation}

\noindent where
\begin{equation}  \label{kappaa}
K=\sqrt{E^{2}-m^{2}(1+\lambda_{r})}\, \qquad \lambda=m^{2}\sqrt{%
\lambda_{r}^{2}-\lambda_{0}^{2}}\,.
\end{equation}

\noindent Considering $F_{0}(0)=0$ and $\int_{0}^{\infty}dr|F_{0}|^{2}<\infty
$, the solution for (\ref{eqoh}) with $K$ and $\lambda$ real is the
well-known solution of the Schr\"{o}dinger equation for the
three-dimensional harmonic oscillator. Note that the condition $\lambda$
real implies that $|\lambda_{r}|>|\lambda_{0}|$, meaning that the radial
component of the nonminimal vectorial potential must be stronger that its
time component in order to the effective potential be a true confining
potential. On the other hand, if $\lambda_{r}=0$ or $|\lambda_{r}|<|%
\lambda_{0}|$, we obtain $\lambda=i|\lambda|$ and the effective potential in
this case will be an inverted harmonic oscillator and the energy spectrum
will consist of a continuum corresponding to unbound states. Therefore, the
presence of radial component of the nonminimal vector potential is an
essential ingredient for confinement.

An detailed study of this effective potential is done in \cite{ben}. Using
the results of Ref.~\cite{ben} the solution is expressed as
\begin{equation}  \label{ener}
|E|=m\sqrt{1+\lambda_{r}+(2n+3)\sqrt{\lambda_{r}^{2}-\lambda_{0}^{2}}}\,
\qquad n=0,1,2,\ldots
\end{equation}
\begin{equation}  \label{funco}
F_{0}(r)=N_{n\,j}r^{j}e^{-\lambda r^{2}/2}L_{\frac{n-j}{2}%
}^{(j+1/2)}(\lambda r^{2})
\end{equation}

\noindent where $N_{n\,j}$ is a normalization constant, $n=2N+j$ with $N$ a
nonnegative integer. Note that $j$ can take values $0,2,\ldots,n$ when $n$
is an even number, and $1,3,\ldots,n$ when $n$ is an odd number and also
that for each value of $j$ there are $2j+1$ different values of $m_{j}$. All
the energy levels are degenerate with the exception of $n=0$. The degeneracy
of the level of energy for a given principal quantum number $n$ is given by $%
(n+1)(n+2)/2$ as a consequence of the presence of essential and accidental
degeneracies.

From (\ref{ener}), we can see that there is an infinite set of discrete
energies (symmetrical about $E=0$) irrespective to sign of $\lambda_{0}$ and
although positive- and negative- energy levels do not touch, they can be
very close to each other for moderately strong coupling constants without
any danger of reaching the conditions for Klein\`{}s paradox. The absence of
Klein\`{}s paradox for this kind of interaction is attributes to fact that
the nonminimal vectorial interaction does not distinguish particles from
antiparticles \cite{ccc2}.

For the case $A_{0}=0$ ($\lambda_{0}=0$), the solution is expressed as
\begin{equation}  \label{enerA0}
|E|=m\sqrt{1+\lambda_{r}+(2n+3)|\lambda_{r}|}\,
\qquad n=0,1,2,\ldots
\end{equation}
\begin{equation}  \label{funcoA0}
F_{0}(r)=N_{n\,j}r^{j}e^{-\Omega r^{2}/2}L_{\frac{n-j}{2}%
}^{(j+1/2)}(\Omega r^{2})
\end{equation}

\noindent where $\Omega=m^{2}|\lambda_{r}|$.

\subsubsection{$(-1)^{j+1}$ parity states}

As mentioned in the section \ref{sec322}, we can not consider the general case (\ref{potl}), because we are not able to obtain analytical solutions to this kind of parity states.

Otherwise, considering (\ref{potl}) with $A_{0}=0$ ($\lambda_{0}=0$) and using the notation, $\Phi_{+}=H_{0}$ and $\Phi_{-}=\phi$, the equations (\ref{schro2}) and (\ref{schro3}) reduce to
\begin{equation}\label{e2oh0}
\frac{d^{2}\Phi_{\pm}}{dr^{2}}+\left[ K_{\pm}^{2}-\Omega^{2}r^{2}-\frac{%
j(j+1)}{r^{2}} \right]\Phi_{\pm}=0\,,
\end{equation}

\noindent where
\begin{equation}\label{kplus}
    K_{+}=\sqrt{E^2-m^{2}(1-\lambda_{r})}\,,
\end{equation}
\begin{equation}\label{kminus}
    K_{-}=\sqrt{E^{2}-m^{2}(1-2\lambda_{r})}\,.
\end{equation}

\noindent The solution for (\ref{e2oh0}) with $K_{\pm}$ and $\Omega$ real is the
solution of the Schr\"{o}dinger equation for the
three-dimensional harmonic oscillator, as in the case of states of $(-1)^{j}$ parity.

The energy can be obtained from the relation
\begin{equation}\label{vinenergy}
    K_{\pm}^{2}=(2n_{\pm}+3)\Omega\,.
\end{equation}

\noindent Now we move on to match a common energy to spin-one particles problem for states of $(-1)^{j+1}$ parity. The compatibility of the solutions for $\Phi_{+}$ and $\Phi_{-}$ demands that the quantum number $n_{+}$ and $n_{-}$ must satisfy the relation
\begin{equation}\label{nplunminus}
    n_{-}-n_{+}=\frac{1}{2}\frac{\lambda_{r}}{|\lambda_{r}|}\,.
\end{equation}

\section{Final remarks}

\bigskip In this Review Article, we showed the correct use and also
presented a few properties of the nonminimal vector interactions in the
Duffin-Kemmer-Petiau (DKP) formalism. A relativistic wave equation must
carry a conserved four-current to exhibit symmetries in physical problems.
In this spirit, we showed that the four-current is not conserved when one
uses either the matrix potential $P\beta^{\mu}$ or $\beta^{\mu}P$ (widely
used in the literature), even though the linear forms constructed from those
matrices potentials behave as true Lorentz vectors. Also, we presented the
correct equations for the problem addressed in \cite{erro}. In this case, we
found an equation very similar to DKP equation for spin-zero particles in
(3+1) dimensions, except for some additional terms. Therefore, the solution
of the three-dimensional DKP equation with nonminimal vectorial interactions
can be found by solving Schr\"{o}dinger-like equations. As an application
of the procedure developed, we considered the problem of spin-one particles
in the presence of a nonminimal linear vector potential and discussed the
necessary conditions in order to the effective potential to be true
confining potential. The absence of Klein\`{}s paradox is attributes to fact
that the nonminimal vectorial interaction does not distinguish particles
from antiparticles \cite{ccc2}.

Ours results are definitely useful because they shed some light on the
understanding of the nonminimal vector interactions. Furthermore, the
correct use of the nonminimal vectorial interaction may be useful due to
wide applications in the description of elastic meson-nucleus scattering.

\bigskip\bigskip \noindent \textbf{Acknowledgments}
\bigskip

The authors are indebted to the anonymous referee for an excellent and
constructive review. This work was supported in part by means of funds
provided by CAPES. This work was partially done during a visit (L. B.
Castro) to UNESP - Campus de Guaratin\-gue\-t\'{a}.

\bigskip\bigskip

\noindent \textbf{Appendix A: The vector spherical harmonics}

\bigskip \noindent The properties of the vector spherical harmonics used in
this work are obtained from \cite{hill}. The list of properties is the
following
\begin{equation}  \label{apb1}
\hat{r}Y_{j}^{m_{j}}=-\alpha_{j}\vec{Y}_{j\,j+1\,m_{j}}+\zeta_{j}\vec{Y}%
_{j\,j-1\,m_{j}}
\end{equation}
\begin{equation}  \label{apb2}
\vec{\nabla}Y_{j}^{m_{j}}=\alpha_{j}\frac{j}{r}\vec{Y}_{j\,j+1\,m_{j}}+%
\zeta_{j}\frac{j+1}{r}\vec{Y}_{j\,j-1\,m_{j}}
\end{equation}
\begin{equation}  \label{apb3}
\vec{\nabla}\cdot(f(r)\vec{Y}_{j\,j+1\,m_{j}})=-\alpha_{j}\left( \frac{df}{dr%
}+\frac{j+2}{r}f \right)Y_{j}^{m_{j}}
\end{equation}
\begin{equation}  \label{apb4}
\vec{\nabla}\cdot(f(r)\vec{Y}_{j\,j\,m_{j}}) =0
\end{equation}
\begin{equation}  \label{apb5}
\vec{\nabla}\cdot(f(r)\vec{Y}_{j\,j-1\,m_{j}})=\zeta_{j}\left( \frac{df}{dr}-%
\frac{j-1}{r}f \right)Y_{j}^{m_{j}}
\end{equation}
\begin{equation}  \label{apb6}
\vec{\nabla}\times(f(r)\vec{Y}_{j\,j+1\,m_{j}})=i\zeta_{j}\left( \frac{df}{dr%
}+\frac{j+2}{r}f \right)\vec{Y}_{j\,j\,m_{j}}
\end{equation}
\begin{equation}  \label{apb7}
\vec{\nabla}\times(f(r)\vec{Y}_{j\,j\,m_{j}})=i\zeta_{j}\left( \frac{df}{dr}-%
\frac{j}{r}f \right)\vec{Y}_{j\,j+1\,m_{j}} +i\alpha_{j}\left( \frac{df}{dr}+%
\frac{j+1}{r}f \right)\vec{Y}_{j\,j-1\,m_{j}}
\end{equation}
\begin{equation}  \label{apb8}
\vec{\nabla}\times(f(r)\vec{Y}_{j\,j-1\,m_{j}})=i\alpha_{j}\left( \frac{df}{%
dr}-\frac{j-1}{r}f \right)\vec{Y}_{j\,j\,m_{j}}
\end{equation}
\begin{equation}  \label{apb9}
\hat{r}\times\vec{Y}_{j\,j+1\,m_{j}}=i\zeta_{j}\vec{Y}_{j\,j\,m_{j}}
\end{equation}
\begin{equation}  \label{apb10}
\hat{r}\times\vec{Y}_{j\,j\,m_{j}}=i\zeta_{j}\vec{Y}_{j\,j+1\,m_{j}}+i%
\alpha_{j}\vec{Y}_{j\,j-1\,m_{j}}
\end{equation}
\begin{equation}  \label{apb11}
\hat{r}\times\vec{Y}_{j\,j-1\,m_{j}}=i\alpha_{j}\vec{Y}_{j\,j\,m_{j}}
\end{equation}
\begin{equation}  \label{apb12}
\hat{r}\cdot\vec{Y}_{j\,j+1\,m_{j}}=-\alpha_{j}Y_{j}^{m_{j}}
\end{equation}
\begin{equation}  \label{apb13}
\hat{r}\cdot\vec{Y}_{j\,j\,m_{j}}=0
\end{equation}
\begin{equation}  \label{apb14}
\hat{r}\cdot\vec{Y}_{j\,j-1\,m_{j}}=\zeta_{j}Y_{j}^{m_{j}}
\end{equation}

\end{document}